\documentclass[12pt,epsf]{article}
\usepackage{epsfig}

\newcommand{\onefigure}[2]{\begin{figure}[htbp]
\begin{center}\leavevmode\epsfbox{#1.eps}\end{center}\caption{#2\label{#1}}
\end{figure}}

\renewcommand{\thanks}[1]{\footnote{#1}} % Use this for footnotes

\newcommand{\be}{\begin{equation}}
\newcommand{\ee}{\end{equation}}
\newcommand{\bea}{\begin{eqnarray}}
\newcommand{\eea}{\end{eqnarray}}

\begin{document}

\begin{flushright}
SLAC-PUB-11063\\
18 March 2005\\
\end{flushright}

\bigskip\bigskip

\begin{center}
{\bf\large SCIENTIFIC ESCHATOLOGY\footnote{\baselineskip=12pt Work
supported in part by Department of Energy contract
DE--AC03--76SF00515.}}
\end{center}

\begin{center}
H. Pierre Noyes and James V. Lindesay\footnote{Permanent address:
Department of Physics and Computational Physics Lab, Howard
University,
Washington, D.C. 20059, jlslac@slac.stanford.edu} \\
Stanford Linear Accelerator Center MS 81,\\
Stanford University, 2575 Sand Hill Road, Menlo Park CA 94025,
noyes@slac.stanford.edu\\
\end{center}

\begin{center}
{\bf Abstract}
\end{center}
The future evolution of the universe suggested by the cosmological
model proposed earlier at this meeting by the authors is explored.
The fundamental role played by the positive ``cosmological
constant" is emphasized. Dyson's 1979 paper entitled \emph{Time
Without End} is briefly reviewed. His most optimistic scenario
requires that the universe be \emph{geometrically} open and that
biology is structural in the sense that the current complexity of
human society can be reproduced by scaling up its (quantum
mechanical) structure to arbitrary size. If the recently measured
``cosmological constant" is indeed a fundamental constant of
nature, then Dyson's scenario is, for various reasons, ruled out
by the finite (De Sitter) horizon due to exponential expansion of
the resulting space. However, the finite temperature of that
horizon does open other interesting options. If, as is suggested
by the cosmology under consideration, the current exponential
expansion of the universe is due to a phase transition which fixes
a \emph{physical} boundary condition during the early radiation
dominated era, the behavior of the universe after the relevant
scale factor crosses the De Sitter radius opens up still other
possibilities. The relevance of Martin Rees' apocalyptic
eschatology recently presented in his book \emph{Our Final Hour}
is mentioned. It is concluded that even for the far future,
whether or not cultural and scientific descendants of the current
epoch will play a role in it, an understanding (sadly, currently
lacking) of community and political evolution and control is
essential for a preliminary treatment of what could be even
vaguely called \emph{scientific} eschatology.

\bigskip

\begin{center}
A preliminary version of this paper was presented at\\
The Twenty-Sixth Annual Meeting of the\\
{\bf ALTERNATIVE NATURAL PHILOSOPHY ASSOCIATION}\\
Cambridge, England, 31 July - 5 August, 2004\\
\end{center}

\bigskip

\section{Introduction}

On the Silver Jubilee of the founding of the Alternative Natural
Philosophy Association it was possible to present a new piece of
natural philosophy\cite{SciCos04, LNsub04, LJNiprep04} which the
authors believe has profound implications for the new field of
particle-astrophysics/cosmology; SLAC has recently made this
discipline an important part of its mission. What follows is a
discussion of a few of the important questions raised by the
future implications of this cosmology if it, and other cosmologies
which share with it a positive ``cosmological constant", survive
the rigorous scientific scrutiny they are now receiving. The
authors believe that the way some of these implications are
absorbed by our science and the broader cultures in which our
science is embedded will have relevance to the enormous problems
the children and grandchildren of those present here will have to
face in the coming decades.

During the last fifty years, scientific cosmology has moved from
being a speculative field which could be viewed askance by many
scientists not intimately associated with it to being one of the
``cutting edge" disciplines --- both theoretically and
observationally --- of 21st century science. But most of the
technical work in cosmology is confined, necessarily, to the study
of the past. The study of the future of the universe rarely gets
such careful attention. One goal of this paper is to try to
motivate more working scientists to bring about a similar change
of attitude and practice in what we call here \emph{scientific
eschatology}.

Coincidentally with the founding of ANPA, Dyson\cite{Dyson79}
published a paper with the same objective. At that time he could
take a geometric view of the problem, defined by whether the
universal curvature parameter is positive (closed universes) or
negative (open universes). One of his objectives in the paper was,
clearly, to make a case for the possibility that rational,
scientific, ecological communities of living organisms can look
forward to a ``time without end". He easily shows that if the
members of such communities depend on the biochemistry with which
we are familiar, or, in more colloquial terms, contain ``flesh and
blood" organisms as an essential constituent, there is no hope
that his objective can be reached within either class of
cosmologies he considers. He therefore extends the definition of
``biology" to include ``organisms" (and by implication, ecosystems
composed of them) that have the same \emph{structure} as those we
have become familiar with on our planet. This was done by
postulating a scaling law for quantum mechanical interactions
which gave precision to his meaning of \emph{structure}. Even in
this context, he was unable to meet his objective of preserving
rationality forever within closed (``big crunch") cosmologies. But
for open cosmologies he found a way to show that appropriate
``biological" strategies could provide a (subjective) \emph{time
without end} for organisms and ecosystems of ever increasing
complexity. We will summarize his arguments below.

At the time Dyson wrote, observational evidence for a
``cosmological constant" did not exist. Einstein's motivation for
introducing this constant (i.e. to preserve, in the large, a
static and infinite universe satisfying his \emph{cosmological
principle}) had evaporated with discovery of the red shift for
distant galaxies and Hubble's law. So there was no reason for
Dyson to include in his discussion the De Sitter cosmologies
produced by a positive cosmological constant. We now know that
until about 5 Giga-years ago the rate of expansion of the universe
was \emph{decreasing}.  We also know that more recently the rate
has been \emph{increasing}. From very early times until the
present ($\sim 13.6 \ Gyr$) the observational data can be fitted
by a positive cosmological constant $\Lambda$ together with an
evolving matter density which can be checked in other ways. The
data are consistent with $\Lambda$ being strictly constant
throughout this period. If $\Lambda$ is indeed a ``constant of
nature", this fact would render Dyson's analysis moot.

A review of Dyson's paper presented here reaches the conclusion
that, under his assumptions (which did not include the possibility
of a cosmological constant), a ``time without end" for the type of
``biology" he defines was then a live possibility. Next, the
consequences of having a cosmological constant fixed for all time
will be examined. At first sight, these consequences are indeed
dismal. They could be met by the tragic remarks of Bertrand
Russell\cite{Russell1905} in \emph{A Free Man's Worship} or the
defiant challenge with which Dyson closes his own article:
\begin{quotation}
As Haldane (1924) \cite{Haldane24} the biologist wrote some fifty
years ago,  ``The human intellect is feeble, and there are times
when it does not assert the infinity of its claims. But even then:
\begin{center}
Though in black jest it bows and nods\\
I know it is roaring at the gods\\
Waiting the last eclipse,"
\end{center}
\end{quotation}

Fortunately, we believe there are possible routes to ``time
without end" which might be reached from where we are in an
accelerating expansion scenario. The physics in the class of
models presented earlier in this meeting\cite{SciCos04} need not
\emph{require} the ``cosmological constant" to be fixed forever,
--- only over a well defined and finite period of time. This opens
up a new set of possibilities for biological strategies, which
might allow our intellectual and cultural heirs to win through to
an indefinitely extendable future. The authors of this paper have
only scratched the surface of this fascinating subject, but this
analysis suggests that heroic action may well be required, not
just now, but for many Giga-years into the future. That prompt
action in this century is clearly required has been recently
argued by the Astronomer Royal\cite{Rees04}. Older members of ANPA
will recall, and more recent members should be made aware, that
Martin Rees played a highly significant role in supporting the
first tottering steps of ANPA toward the robust success we now
enjoy. The authors of this article trust he will not mind too much
if we take his measured analysis of the current desperate
situation as the basis for a plea to the scientific community at
large to recognize the tremendous need for a rigorous
\emph{political} science; this argument clearly goes beyond the
case Dyson has made for including an extended scientific
\emph{biology} in order to establish an adequate basis for
\emph{scientific eschatology}. But to make this case we must now
complete the steps we have already indicated that lead us to that
conclusion.

\section{Dyson's ``Time without end"}

We now take a closer look at Dyson's pioneering paper. As already
noted he rapidly concludes that a ``big crunch" universe does not
provide enough scope for an optimistic view of the far future.
Even in an open (or flat) universe, Dyson's argument closes the
door on the continuation of ``flesh and blood" biology  of a
complexity comparable to that currently experienced on our planet
for a \emph{physical} time without end. This focuses his interest
on the biology of ``...sentient black clouds\cite{Hoyle57}, or
sentient computers\cite{Capek23}...",--- structures which we know
some (most?) ANPA members dismiss as impossible. To make his
argument plausible and quantitative, he proposes a ``biological
scaling hypothesis" that has as its first consequence the
conclusion that the ``...appropriate measure of time as
experienced subjectively by a living creature is not physical time
\emph{t} but ..." an integral from zero to \emph{t} of the
temperature function defined by his scaling law (\cite{Dyson79},
Eq.56).  This is called \emph{subjective} time. ``The second
consequence of the scaling law is that any creature is
characterized by a quantity \emph{Q} which measures its rate of
entropy production per unit of subjective time."  For a human
being living long enough to pronounce ``Cogito, Ergo Sum" this
works out to be $\sim 10^{23}$ bits, and for our species $\sim
10^{33}$ bits. This sets a fixed lower bound for the temperature
$\theta $, which is (\cite{Dyson79}, Eq. 73) $\theta > (Q/N) \
10^{-12} deg$, where \emph{N} is the number of electrons available
to the society of complexity \emph{Q}. For our current biosphere
$N=10^{42}$, so our current social complexity cannot be maintained
at temperatures lower that $10^{-23} deq$.

Here a few comments are in order. Dyson's biological scaling
hypothesis (\cite{Dyson79}, p. 454) is

\begin{quotation}

\emph{Biological Scaling Hypothesis.  If we copy a living
creature, quantum state by quantum state, so that the Hamiltonian
of the copy is
\[ H_c = \lambda UHU^{-1} \, ,
\]
where $H$ is the Hamiltonian of the creature, $U$ is a unitary
operator, and $\lambda$ is a positive scaling factor, and if the
environment is similarly copied so that the temperatures of the
environments of the creature and the copy are respectively $T$ and
$\lambda T$, then the copy is alive, subjectively identical to the
original creature, with all its vital functions reduced in speed
by the same factor $\lambda$.}

\end{quotation}

This hypothesis is made in a context which should be spelled out
further. Dyson starts his section on biology by posing three deep
questions concerning the nature of life and consciousness:

\begin{quotation}

$ \ \ $ (i) Is the basis of consciousness matter or structure?

$ \ \ $ (ii) Are sentient black clouds, or sentient computers,
possible?

$ \ \ $ (iii) Can we apply scaling laws in biology?

\end{quotation}

As is clear in his article, if the answer to (i) is ``matter",
Dyson takes this as, in more colloquial language, tying
consciousness inexorably to ``flesh and blood" and in the
cosmological context to certain death. There are traditional ways
(eg religious rather than scientific) to escape this dismal
conclusion, but the authors of this article follow, instead,
Dyson's optimistic attitude and assume that the basis of
consciousness is structural rather than material. The reader
should consult his article (and other writings) to see how he
justifies his own attitude. He notes that we (i.e. our culture) do
not yet know how to answer these three questions ``But they are
not in principle unanswerable. It is possible that they will be
answered fairly soon as a result of progress in experimental
biology."

As just explained, Dyson's tentative answer to question (i) is
``structure". The fact that ``quantum computers" are now claimed
to have greatly extended powers compared to ``classical computers"
lends weight to his tentative conclusion.  His tentative ``yes"
answer to (ii) puts him in the ``strong AI" camp. The authors of
this article are not sure this is the right (even tentative)
answer today, but HPN feels it likely  that \emph{communities} of
computers starting in environments with appropriate resources and
with inheritance mechanisms and survival pressures sufficiently
similar to those encountered in biology would evolve into
recognizably conscious beings. The technical point here is that a
community that has evolved to the point where its members have
independent choices of action available (i.e. have ``free will")
is obviously not an \emph{algorithmic computer}.

Support for Dyson's tentative ``yes'' answer to (iii) is provided
by recent work by Fred Young\cite{Young05}, a former president of
ANPA. Young makes a strong case that the basic structures of
biology can be represented by scaling laws using small-protein
concentration ratios for physiological modelling at the cellular
level, tissue and system organizational modelling at the organism
level, biological species organization at the ecological level,
etc. The authors of this article find Young's results to date
quite compelling and extremely promising for the future of his
approach to biology.

We now return to Dyson's paper at the point where he has
established the finite temperature bound $\theta > (Q/N) \
10^{-12} deg$ below which any society of complexity \emph{Q} bits
controlling \emph{N} electrons must not fall. This bound is
arrived at by asserting that the rate of energy dissipation (i.e.
use of energy by the society) must not exceed the power that can
be radiated away into space.  At that point it can still dissipate
the energy which it must expend to keep on operating. Below that
point it must decrease its complexity or increase its controlled
number of electrons. Since the supply of energy available to the
society is assumed finite, it must reach this point at a finite
time, and Dyson remarks (\cite{Dyson79}, p. 456):

\begin{quotation}

We have reached the sad conclusion that the slowing down of
metabolism described by my biological scaling hypothesis is
insufficient to allow a society to continue indefinitely.

\end{quotation}

Here Dyson examines a possible way to insure ``time without end"
for entities who ``live" in this cold and forbidding future. This
is, quite simply, the biological strategy of \emph{hibernation}.
Life can metabolize at a higher temperature and then hibernate at
a much lower temperature to stretch out its \emph{subjective}
time. To quote Dyson again (\cite{Dyson79}, p 4550])

\begin{quotation}

Suppose then that a society spends a fraction $g(t)$ of its time
in its active phase and a fraction $[1-g(t)]$ hibernating. The
cycles of activity and hibernation should be short enough so that
$g(t)$ and $\theta (t)$ [Here $\theta (t)$ is a function Dyson has
already assumed technologically available subject to explicit
thermodynamic constraints] do not vary appreciably during any one
cycle. Then (56) and (59) [previous constraints] no longer hold.
Instead subjective time [$u(t)$] is given by
\begin{center}
$u(t) = f \int_0^t g(t') \theta (t')dt'$,
\end{center}
[rather than $u(t) = f \int_0^t \theta (t')dt'$]

\end{quotation}
and is no longer bounded. The parameter $f$ was chosen by Dyson to
have a value of $(300 \, deg \, sec)^{-1}$ as a scale comparable
to that of human society to make the subjective time rate $u(t)$
dimensionless. Hence the constraints which led to his dismal
conclusion no longer applied. In this way he was able to achieve a
system with an infinite subjective time in an expanding universe,
while expending only a finite amount of energy.

One matter where the finite energy limitation is serious is in the
storage of memory. As Dyson remarks (\cite{Dyson79}, p. 456)

\begin{quotation}

I would like our descendants to be endowed not only with an
infinitely long subjective lifetime but also with a memory of
endlessly growing capacity. To be immortal with a finite memory is
highly unsatisfactory; it seems hardly worth while to be immortal
if one must erase all trace of one's origins in order to make room
for new experience.

\end{quotation}

Digital memory with the finite energy resources available to a
periodically  hybernating society is obviously not an option which
meets this requirement when {\it digital} memory storage in
condensed matter is employed. Dyson turns to analog memory and
claims that the angles between a finite number of structures (e.g.
stars) in an expanding universe can be used for an \emph{analog}
memory storage of ever increasing capacity.  So far as we can see
here, the point to focus on is that for indefinitely expanding
storage capacity for \emph{information} to be available, the
system must have no finite upper bound on the {\it entropy}. In
Dyson's expanding universe scenario, even though the energy
available to the society is finite, the unbounded expansion of the
volume over which this energy is distributed means that there is,
indeed, no {\it a priori} upper limit on the entropy. Consequently
analog storage might meet the requirement he imposes. As with much
of this discussion, meeting {\it technological} challenges this
requirement poses starting from any particular configuration could
prove to be daunting.

\section{Asymptotically De Sitter Universes}

Recent observational results have tentatively convinced most of
the experts that the energy density of our universe is currently
partitioned into approximately 73\%  dark energy, 23\% dark matter
and 4\% ordinary matter and radiation, in a space that is flat
rather than either ``closed" or ``open". The simplest way to fit
the data is to assume that the various interlocking pieces of
evidence which lead to this picture constitute an actual
\emph{discovery} of Einstein's cosmological constant $\Lambda$ as
a new universal constant \emph{and} a measurement of that constant
to a couple of percent. In the approach taken in the paper
presented earlier in this meeting\cite{SciCos04}, these authors
prefer to think of it as a \emph{phenomenological} constant
specified as constant only over a finite interval in universal
time. That point of view is explored eschatologically in the next
section. In this section we adopt the more naive approach.

The specific consequences of interest here which follow from this
assumption are that:

a) The matter energy density that drives the cosmological
expansion in the dynamical Friedman-Lemaitre (FL) equation, which
we call $\rho_{FL}$, will eventually become insignificant compared
to the cosmological constant density $\rho_{\Lambda} = {\Lambda
c^4\over 8\pi G_N}$ (here $G_N$ is Newton's gravitational
constant).

(b) Consequently, the FL (Hubble) equation is replaced
asymptotically by $[{\dot R\over R}]^2 \simeq {8\pi G_N\over 3c^2}
\rho_{\Lambda} ={\Lambda c^2/3}$, which implies that $R(t)=R_E
e^{{\sqrt {\Lambda \over 3}}ct} \equiv R_E e^{{ct\over
R_{\Lambda}}}$. Here $R_{\Lambda}$ is sometimes called the De
Sitter radius or horizon, and $ R_E$ is set at a time when
$\rho_{FL}$ is negligible compared to $\rho_{\Lambda}$. Although
objects of the scale of the gravitationally bound super-cluster
have dynamics which are determined by local matter densities, the
late time exponential expansion defines a cosmological De Sitter
horizon $R_{\Lambda} = \sqrt{{3\over\Lambda}} \simeq 16.6 \ Glyr$
which serves as a causal boundary, i.e. anything which crosses
this boundary can never re-establish luminal contact with our
region of the universe. Our galaxy appears to be close to the edge
of, and probably bound to a super-cluster with a radius of about
50 mega-parsecs $\approx 0.16 \, Gly$. If we therefore take the
current FL scale parameter ($R_0= R(t_0)$ at time $t_0$) in our
(gravitationally bound) locality to be about 100 mega-parsecs, we
find that this scale parameter will cross the De Sitter horizon
(i.e. $R(t) = R_{\Lambda}$) when $t \simeq t_0 + 65 Gyr$. Here the
current time $t_0$ is usually taken to be about $13.6 \ Gyr$. This
means that all other galaxies not in our local super-cluster will
vanish within about $65 \ Gyr$. This consequence already precludes
useful discussion of Dyson's analog memory storage and far-ranging
communications strategies if $\Lambda$ is indeed a universal
constant on a par with $\hbar, c, G_N $ and $k_B$. Both of these
strategies rely on scaling laws that assume causal (i.e. luminal)
contact can always exist if the society waits long enough. For
further discussion see (c), (i) and Section 4 below.

(c) Since the De Sitter horizon has a finite area, the causal
region of the De Sitter space will have\cite{LSJL05} a finite
entropy $S_\Lambda =k_B{\pi R_{\Lambda}^2\over L_P^2}$, where
$L_P=(\hbar G_N)^{{1\over 2}}c^{-{3\over 2}}$ is the Planck length
and $k_B$ is Boltzmann's constant.
\begin{quote}
 (i) Because finite entropy implies finite information storage
 capacity, this fact precludes, simply using counting arguments, any
 way of realizing Dyson's analog storage method for constructing
 an indefinitely extendable memory. As a counting argument, this
 holds for quantum-coherent systems (eg quantum computers) as well
 as for digital computers.

 (ii) Systems with finite entropy undergo Poincar\'{e}
 recurrences\cite{LDJLLS02}. Such recurrences are due to the finite number
 of configurations (microstates) available to a system with finite entropy.
 Because there are only a finite number of configurations that the system
 fluctuates among, the system will eventually return to any given initial
 configuration.  Since these recurrences have only to
 do with counting of states, this fact applies to both classical
 and quantum  statistical systems. Such recurrences are
 maximally destructive of information.

(iii) One strategy of despair that has sometimes been suggested is
to find a way to pass clues about our experience in this universe
through its fiery destruction to provide information that could
prove useful to new societies evolving in the cycle that might
emerge after we are consumed. Clearly, the destruction of
information precludes placing any hope in this possibility.

(iv) This list of unpleasant facts about a De Sitter universe
could easily be extended. This might explain why Dyson has not, to
our knowledge, extended his analysis to such universes.
\end{quote}

There may still be a finite strategy arising from the fact that in
such universes the De Sitter horizon maintains a finite
temperature. The existence of a horizon creates an information
deficit within the space bounded by that horizon.  The subtle
quantum correlations in states near the horizon must be described
statistically in terms of the degrees of freedom and parameters
accessible in the causal region. Whenever there are such
statistical degeneracies in ways to describe a particular physical
state, the concept of temperature becomes a meaningful tool to
describe macroscopic states.  There are many states across any
horizon which can describe any given measured state within the
causal region, thus associating an entropy and temperature with
that horizon. Consequently, this could serve as an inexhaustible
source of energy to run a \emph{steady state} (constant rate of
energy throughput) forever. For instance a sufficiently clever
technological society could focus the thermal energy coming from
some finite area of the horizon on a ``boiler" which could be
maintained at a higher temperature than the horizon. That society
could then employ a thermodynamically viable engine to extract
useful work (e.g. turn heat into low temperature matter or energy
density) by a cycle between the ``boiler" and a ``condenser" at
some temperature intermediate between the boiler temperature and
the horizon temperature. To be viable, the time scale of the work
cycle must be much less than the equilibration time of the heat
engine\cite{Etter05}. This engine will, of course, be a
non-equilibrium element of a much larger thermodynamic system
which includes the De Sitter horizon. The condenser would require
an efficient radiator (essentially already assumed possible by
Dyson) to get rid of the thermodynamically required waste heat for
the cycle. So far as we can see, such a device is not in conflict
with the second law of thermodynamics. This society would have the
inescapable informational problems arising from finite memory, and
political problems arising from having to decide how the finite
resources are budgeted between current and future needs. However,
we will have to find solutions to the political problems if we are
to get through the next century, as we discuss later.

A more ambitious, and more speculative possibility is that the
``steady state" societies we have considered so far are not energy
limited. Consequently, they might even sequester more energy than
they need to keep going, and become energy \emph{accumulating}
societies containing expanding resources of condensed matter, part
of which could be devoted to expanding local resources. Perhaps it
might be possible to use accumulated resources to modify the
cosmology itself.

To explore this possibility, examine Einstein's equation, which
describes the connection of the local geometry to the local energy
densities: \be G_{\mu \nu} \: = \: {8 \pi G_N \over c^4} T_{\mu
\nu} + \Lambda g_{\mu \nu}. \ee We assume that the local,
gravitationally bound, energy densities which remain have
clustered in a spherically symmetric manner.  The space-time
metric for a system with spherical symmetry has the form
$ds^2=g_{tt}c^2 dt^2 + g_{rr}dr^2+ g_{\theta
\theta}d\theta^2+g_{\phi \phi}d\phi^2$.  Upon inserting this form
into Einstein's equation, the radial component of the metric can
be shown to be given by \be \left ( g_{rr} \right )^{-1} \: = \: 1
\, - \, {8 \pi G_N \over c^4 \, r} \int _0 ^ r ( \rho (\tilde{r})
+ \rho_\Lambda ) \tilde{r} ^2 d \tilde{r} . \ee The time-time
component $g_{tt}$ of the metric and the inverse of the space
component $g_{rr}^{-1}$ will have the same zero, which determines
the location of the horizon, and fixes the center as the same as
the center of the mass distribution.

For example, if there is a mass $M$ associated with the galactic
super cluster more or less localized near the center of the
region, the horizon scale $r_H$ associated with the local
cosmology satisfies \be 0 \: = \: 1 - {2 G_N M \over c^2 r_H} -
{r_H ^2 \over R_\Lambda ^2} \: = \: 1 - {R_M \over r_H} - {r_H ^2
\over R_\Lambda ^2} , \ee where $R_M$ is the Schwarzschild radius
$R_M \equiv {2 G_N M \over c^2}$.  Note that if $M=0$, the horizon
is located at $R_\Lambda$ as expected.  It is of interest to note
that if $R_M>{2 R_\Lambda \over 3 \sqrt{3}}$, there is no horizon
in the causal region.  One might hope to be able to eliminate the
De Sitter horizon by increasing the mass $M$, however the society
would be trapped between two horizons which are drawn together as
the mass $M$ is increased.  This would indeed be a hostile
environment for finite beings.

Alternatively, if the mass density $\rho_M$ is uniformly
distributed throughout the region, the horizon scale satisfies \be
r_H \: = \: \sqrt{\left ( {3 c^4 \over 8 \pi G_N} \right ) {1
\over \rho_m + \rho_\Lambda}}. \ee The entropy associated with
this horizon is given by $S=k_B {\pi r_H ^2 \over L_P ^2},$
whereas the temperature is given by $T={\hbar c \over 2 \pi k_B
r_H}.$ Any societal activity which is capable of increasing the
energy density within the causal region would be expected to
utilize processes that preserve entropy, converting entropy from
the larger horizon (at a cooler temperature) into the lower
horizon entropy (of smaller area and higher temperature) added to
the local entropy densities associated with the increased local
energy densities.  Such adiabatic processes will unfortunately not
change the overall finite entropy, and therefore not prevent
recurrences or solve information limits. However, if such
processes can occur, the society has access to increasing energy
supplies, increasing temperatures, and the associated increasing
rates of subjective time as defined by Dyson.

\subsection{A possible societal strategy}

We see that if a society is able to modify the overall energy
contained within the causal region, then the regional cosmology
can be changed.  It remains to demonstrate whether such activities
can be fruitful even in principle.  Suppose the society disperses
heat engines uniformly throughout the region, and uses the cold
materials produced by those engines to construct other engines.
The associated energy densities would then satisfy \be {d \rho_M
\over dt}= \alpha \rho_M \Rightarrow \rho_M = \rho_{Mo} e^{\alpha
t} \, , \label{rho_rate} \ee where $\alpha$ is determined by the
efficiency of the engines in converting horizon heat into cold
materials.  The initial available density of engines $\rho_{Mo}$
is expected to be a fraction of the galactic super cluster density
relative to the De Sitter horizon. Therefore, the society would
make significant modifications to the De Sitter cosmology in the
region in a time of the order \be t_{modify \: De Sitter} \: = \:
{1 \over \alpha} log \left ( {\rho_\Lambda \over \rho_{Mo}} \right
), \ee as long as this time is significantly less than the
Poincar\'{e} recurrence time.

We will estimate the scale of this time by assuming that the
thermal distribution of particles associated with the De Sitter
horizon is the same as that associated with a black body cavity at
that temperature.  The efficiency of a Carnot engine running
between the optimal temperatures is given by $e={W_{out} \over
Q_{in}}=1-{T_H \over T_{boiler}}<1$. The thermal distribution of
low mass particles in the cavity satisfies \be n(\epsilon)
d\epsilon  = {g \over 2 \pi^2 (\hbar c)^3 } {\epsilon ^2 d
\epsilon \over e^{\epsilon / k_B T} -1 }, \ee where $n(\epsilon)$
is the number of quanta per unit energy per unit volume, and $g$
is the degeneracy of the thermal quanta. We will design the
collectors on the heat engines so that they are efficient at
collecting radiations of wavelengths ( $\lambda_{collected}\leq
\lambda$ ) comparable and smaller than the dimension of the
collector of area $\lambda \times \lambda$. We will assume that
the collector scale is considerably smaller than the horizon
scale, which means that the exponential in the Planck formula is
much larger than 1. Only a fraction $f_Q$ of those quanta in a
region of space will have directions toward the collector, and the
rate at which usable quanta strike the collector can be estimated
to be \be {\bar{\Delta N_{collected}}\over \Delta t} \approx f_Q
\left ( \int _{hc/ \lambda} ^ \infty  n(\epsilon) d \epsilon
\right ) \times c \times Area_{collector}. \ee Since the area of
the collector defines the wavelength $\lambda^2$, the rate of the
collection of quanta is estimated to be \be {\bar{\Delta
N_{collected}}\over \Delta t} \approx f_Q {g \over \pi} \left ( {c
\over R_\Lambda} \right ) \: e^{-(2 \pi)^2 {R_\Lambda \over
\lambda}}. \ee This rate is clearly exponentially small in the
horizon scale compared to the scale size of the heat collector. We
expect that the rate of heat conversion $\alpha$ should be related
to the rate of quantum collection multiplied by the average energy
per quantum collected relative to the rest energy $mc^2$ of each
heat engine, in order to satisfy Eq. \ref{rho_rate}. Thus, an
estimate of the rate of exponential growth in energy density in
the society due to the heat engines is given by \be
\begin{array}{c}
\alpha \sim f_Q {g \over \pi} \left ( {c \over R_\Lambda} \right )
{hc/ \lambda \over mc^2} e^{-(2 \pi)^2 {R_\Lambda \over \lambda}} \\
\approx f_Q 2 g  \left ( {c \over R_\Lambda} \right ) {\lambda_m
\over \lambda} \: e^{-(2 \pi)^2 {R_\Lambda \over \lambda}},
\end{array}
\ee where $\lambda_m$ is the Compton wavelength of the mass of
each heat engine.

We now can compare the time scale required for such heat engines
to have (local) cosmological significance with the Poincar\'{e}
recurrence time scale of that cosmology. Recurrences are expected
to occur stochastically on time scales given by\cite{LDJLLS02} \be
t_{recurrence} \: \cong \: t_{reshuffle} \: e^{S_\Lambda \over
k_B}, \ee where $t_{reshuffle}$ is the typical time scale
associated with the microscopic reshuffling of those
configurations that give rise to the finite entropy $S_\Lambda$.
The configurations are counted by the thermodynamic weight
$\Omega$ in the Boltzmann identification $S=k_B log \Omega$. This
reshuffling time can be estimated to be a fraction $f_{R}$ of the
causal transit time ${R_\Lambda \over c}$ across the De Sitter
patch (causal region), giving recurrence times of the order \be
t_{recurrence} \: \cong \: f_{R} \left ( {R_\Lambda \over c}
\right ) \: e^{\pi \left ( {R_\Lambda \over L_P}  \right ) ^2}.
\ee Comparison of the recurrence time with the energy accumulation
rate of the heat engines given by \be t_{modify \: De Sitter} \:
\approx \:  { \pi \over f_Q \, g \, log \left ( {\rho_\Lambda
\over \rho_{Mo}} \right ) }
 \left ( { R_\Lambda \over c} \right )
{\lambda \over \lambda_m} \: e^{(2 \pi)^2 {R_\Lambda \over
\lambda}} \ee gives design constraints on the size of the heat
engines: \be {\lambda \over \lambda_m} \: e^{(2 \pi)^2 {R_\Lambda
\over \lambda}} < { f_{R} \, f_Q \, g \, log \left ( {\rho_\Lambda
\over \rho_{Mo}} \right )  \over \pi  } \: e^{\pi \left (
{R_\Lambda \over L_P}  \right ) ^2}. \ee Although on the left hand
side of the equation, the macroscopic size of the collector
$\lambda$ is expected to be decades of orders of magnitude larger
than the Compton wavelength associated with the mass of the heat
engine, the exponentiation of the square of the horizon scale
relative to the Planck length on the right hand side of the
equation should give considerable flexibility in the design of a
workable engine.

Since there seem to be no inconsistencies in building engines
which can utilize the heat of the horizon to store energy density,
we can speculate on the consequences of such actions. As the
horizon scale shrinks, the temperature associated with the horizon
increases.  Using Dyson's association of the subjective time rate
with the temperature of the organism/society, any evolving
``biological" organisms will have increasing rates of subjective
time.  However, if the process of energy density storage is indeed
adiabatic, the overall entropy of the local cosmology will remain
fixed, leaving the Poincar\'{e} recurrence time unchanged.  Such a
strategy could indeed increase \emph{relative} subjective time
without bound. In such an environment, Dyson's strategy of
hybernation would not be useful, since it would only serve to slow
down relative subjective time.  No mention is here made of the
formidable engineering tasks involved in constructing heat engines
which function efficiently in increasingly hot spaces. We have
been unable to come up with any \emph{a priori} reasons against
such societal intervention on a cosmological scale.  Note that in
contrast to Dyson's scenario which accommodates biology to a
changing cosmological environment, this society alters both the
cosmology and itself to manipulate its subjective time.

\section{Is the ``Cosmological Constant" Really Constant?}

The scenario presented at this meeting only infers that $\Lambda$
be constant during the finite time period for which the rate of
expansion of the relevant FL scale factor is sub-luminal. This
interval is illustrated for a specific choice of scale
factor\cite{SciCos04} in Fig.\ref{rate}, \onefigure{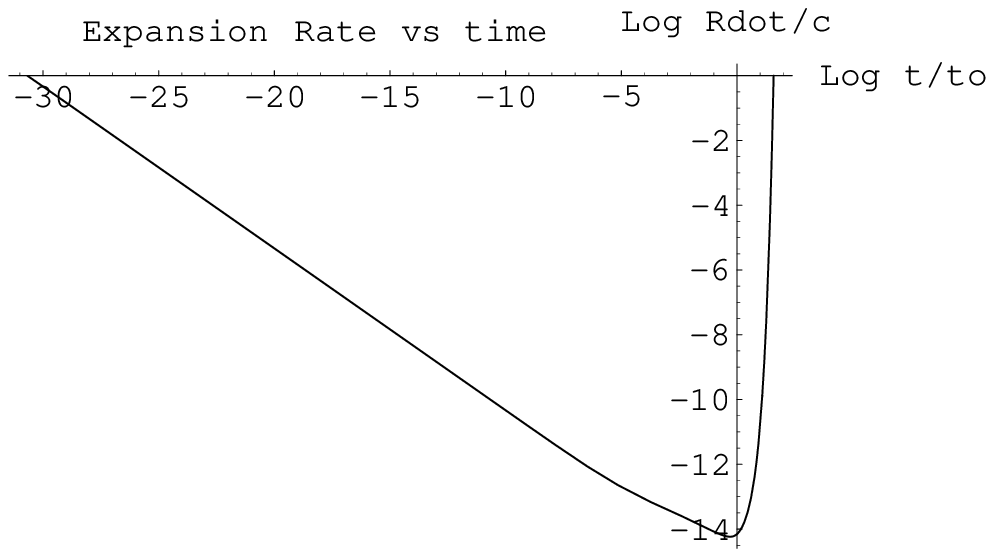}{Finite
time period of sub-luminal expansion rate} but our remarks here
apply to the whole class of models which give a reasonable fit to
current data and meet this requirement. In the model discussed
in\cite{SciCos04}, the earlier time corresponds to a phase
transition characterized by quantum de-coherence such as occurs
when the ground state of a Bose-Einstein condensate starts to
``evaporate" due to the confining density falling below the
critical density at that temperature. Note that in such a model,
density is falling more rapidly than temperature, so that it is
consistent to assume that all of the energy is locked up in the
(quantum-correlated) ground state of the condensate (lowest
spacial frequency mode, using spacially periodic boundary
conditions) \emph{prior} to de-coherence. What happens at the
future time $t_E$ when this scale radius crosses the putative De
Sitter horizon, or even whether there \emph{is} a De Sitter
horizon at that time, is the focus of the discussion in this
section.

To focus our thinking, we will initially assume that the ``dark
energy" which is driving the exponential expansion at late times
(but prior to the second time that $\dot{R}=c$) is ``vacuum
energy" due to zero point motions of sources\cite{Lifshitz}. The
vacuum energy in the Casimir effect depends only on $\hbar c$ and
the boundary conditions, independent of the \emph{coupling
constant} to the electromagnetic field. Nevertheless, it is
possible to calculate the measured force effect using the charges
and currents in the conducting bounding surfaces due to the
fluctuations arising  from the uncertainty principle. This gives
the same result because, physically, the boundary conditions
necessarily require that the boundaries themselves be made of
\emph{material} objects which act as electromagnetic conductors or
dielectrics. Thus, from a \emph{physical} point of view, we might
expect ``dark energy" effects to disappear once the scale radius
of the gravitationally bound portion of the universe we are
considering has crossed the (now putative) De Sitter horizon and
is, \emph{ipso facto}, out of luminal contact.  For instance,
Casimir plates separated by a De Sitter horizon are \emph{not}
expected to exhibit the Casimir effect.  Using the interpretation
of dark energy espoused here, we question whether it will continue
to manifest as a cosmological constant at late times.  The De
Sitter cosmology requires that the cosmological constant be in
fact a constant.  If this is not the case, then expectations and
predictions of a horizon, with its associated properties, are
premature.

At even sooner times, we expect the scale associated with the
cosmological inhomogeneities responsible for galactic clustering
and the fluctuations in the cosmic microwave background radiation
to become comparable to or cross the De Sitter scale radius. As
stated in the previous section, if the expansion is primarily due
to dark energy during the intervening period, this is expected to
occur in about 65 Gyr. Generally, we expect local geometry to be
determined by local energy densities as described using Einstein's
equation $ G_{\mu \nu}(x) \: = \: 8 \pi G_N T_{\mu \nu}(x) +
\Lambda g_{\mu \nu}(x) . $ For dynamically significant periods of
time prior to this crossing, it is clear that  the homogeneity and
isotropy assumptions inherent in a Friedman- Lemaitre cosmology do
not hold on the scale of galactic clustering.  This means that the
local geometry generated, $G_{\mu \nu}(x)$, is neither pure
(cosmological) FRW-Lemaitre nor the Schwarzschild-like region of
an isolated galactic cluster in Minkowski space (which would have
no space-time expansion from dark energy).  For instance, our
local gravity is primarily the Schwarzschild space-time generated
by Earth, with negligible influence from the overall cosmological
acceleration due to the dark energy (or else we would be leaving
the surface of the Earth!).  This means that our local space-time
is not undergoing the exponential expansion associated with a
cosmological constant, despite our presence in an accelerating
cosmology.  We expect the evolution of our local scales to be
determined by our local energy (and dark energy) densities,
appropriately matching asymptotic boundary conditions.  Likewise,
on scales for which the cosmological matter inhomogeneities are
important, the local densities are expected to have significant
influence on the behavior of the geometry relative to cosmological
dynamics.  As the scale of relevance to galactic clustering
crosses the De Sitter scale radius, one must take care in
describing the De Sitter scale as a horizon. It is not
unreasonable to suggest that the association of a given scale
distance with supra-luminal rates of expansion could be only a
temporary phase in the evolution of a cosmology that contains
radiation, matter, and dark energy.

Once all cosmologically ``co-moving"  matter associated with the
current exponential expansion has lost (luminal) causal contact
with the finite, gravitationally bound system we believe will be
left behind, we expect the regional situation to change. One
conjecture is that from then on there is no reason to believe that
the De Sitter scale radius, having no physical system to support
it, should remain a ``horizon" (meaning that regions beyond this
``horizon" which were receding supra-luminally would again move
sub-luminally).  More precisely, the horizon (which is a global
concept) never really existed, but there would only be a temporary
loss of luminal contact as objects cross into a region which will
recess at supra-luminal rates. After that time, objects which
achieve (necessarily sub-luminal) escape velocity from the finite
system (which now will hardly be describable as having ``uniform
density" on scales comparable to the De Sitter radius) would
presumably continue to spread out into the \emph{flat space} which
is then the appropriate boundary condition for describing escape
velocity. They could well be out of reach in terms of intact
recovery as objects. However they would still remain in (eventual)
luminal contact. This conjecture raises eschatalogically important
opportunities and issues which we now explore.

Fortunately, there would be no information horizon limiting the
potential complexity of memory.  This means that the entropy of
the system need no longer be finite, and hence there would be no
``big crunch" due to a Poincar\'{e} recurrence.

Unfortunately there would be no cosmological heat source to draw
upon for energy, so that the finite energy crisis would be
exacerbated.  This implies that in the earliest stages of this
scenario all efforts should be made to collect energy resources to
be utilized during the cold far future. Hibernation would be a
very bad idea until AFTER causal contact begins to be
re-established with the remaining accessible parts of the
universe.

Our position on the fringes of the bound galactic super cluster is
advantageous for the transitional stage of the eschatology.
Regions in the bound cluster nearest the outer orbits are well
positioned with regards to communications, access to external
information, and energy required for transportation.  It looks as
though we are destined to be on the dynamic frontier.

\section{Apocalyptic Eschatology}

Up to now the assumption has been implicitly made that the
eschatological problem of primary concern has been whether
``biology" in  Dyson's sense can continue indefinitely (at least
in subjective time) at the level of complexity our civilization
has already achieved and with an ever growing memory. As has been
seen, the technological challenges are formidable, but nothing in
the laws of physics as now known precludes this possibility with
anything like certainty. Implicit also in this analysis is the
assumption that strict causality, or in theological terms
predestination, does not hold. In other words, such societies are
assumed to have, in some effective sense, \emph{free will}, that
is to make choices which have meaningful consequences relevant to
the survival of themselves and/or their heirs. For better or
worse, we so far only have knowledge of \emph{one} such society
that has passed the technological threshold needed to even
envisage the far stretches of time involved. Hence we have no
\emph{scientific} way to estimate the probability of success. But
we do have available reasonably well understood examples of
societies at somewhat lower levels of technological development
which have not had the foresight to avoid collapse, or even
extinction\cite{CoCS90,Diamond05}.

For a thoughtful analysis of the current situation we turn to the
Astronomer Royal, Martin Rees\cite{Rees04}, who will be remembered
by older ANPA members for his invaluable assistance in getting
ANPA going. He comes to the shocking conclusion that our global
society has only a 50-50 chance of surviving the challenges we
will meet in the current century. What strikes the authors of this
paper as most depressing in the picture he paints is not just the
individual problems --- which are threatening enough --- but the
fact that he, in common with Dyson's earlier treatment of the far
future, fails to discuss the fact that even if a technological
means of meeting the problems is conceivable, there is no
\emph{global} decision making process in prospect, let alone
available, that can bring together the planetary resources needed
and direct them into the search for and implementation of the
action needed on the time scale available. A promising start on
the analysis of the problem of environmental collapse has been
made by Jared Diamond, using a broad enough sample of examples to
be meaningful. He finds, somewhat to his surprise, that there are
no cases of collapse due solely to environmental change
(\cite{Diamond05}, p. 11). Four of the five different sets of
factors needed to make the analysis (environmental damage, climate
change, hostile neighbors, friendly trading partners) can be more
or less important or even absent, but understanding the society's
responses to its environmental problems \emph{invariably} is
needed to understand the result. In other words, the basic problem
falls squarely in the political arena, as our own analysis of the
situation had concluded prior to encountering this recent
development in his work. In short, the \emph{political science}
needed for the task has yet to be created. This is not the place
to suggest how that might be achieved, other than the stale remark
that without such a guide to global mobilization, the future is
bleak.

To end on a more cheerful note, once our species succeeds in
meeting the political problem, and avoiding the threatening
apocalypse, the future for our intellectual and cultural heirs
could well continue as long as the political will to do so
persists.


\begin{thebibliography}{99}
\bibitem{Etter05}
The authors are indebted to Tom Etter for pointing this out.
\end{thebibliography}
\end{document}